\documentclass[final,nofootinbib]{revtex4}

\usepackage{epsfig,graphicx}
\usepackage{amssymb}
\usepackage{amsfonts}
\usepackage{bm}

\begin{document}
\setcounter{table}{0}
\def\thetable{\arabic{table}}
\def\thesection{\arabic{section}}

\title{Dielectron widths of the S-, D-vector bottomonium states}

\author{\firstname{A.~M.}~\surname{Badalian}}
\email{badalian@itep.ru}
\affiliation{Institute of Theoretical and Experimental Physics, Moscow, Russia}

\author{\firstname{B.~L.~G.}~\surname{Bakker}}
\email{blg.bakker@few.vu.nl} \affiliation{Department of Physics
and Astronomy, Vrije Universiteit, Amsterdam, The Netherlands}

\author{\firstname{I.~V.}~\surname{Danilkin}}
\email{danilkin@itep.ru}\affiliation{Moscow Engineering Physics Institute,
Moscow, Russia}
\affiliation{Institute of Theoretical and Experimental Physics, Moscow, Russia}


\begin{abstract}
The dielectron widths of $\Upsilon(nS)~(n=1,\dots,7)$ and vector
decay constants are calculated using  the Relativistic String
Hamiltonian with a universal interaction. For $\Upsilon(nS)~
(n=1,2,3)$ the dielectron widths and their ratios are obtained in
full agreement with the latest CLEO data. For $\Upsilon(10580)$
and $\Upsilon(11020)$ a good agreement with experiment is reached
only if the $4S$---$3D$ mixing (with a mixing angle
$\theta=27^\circ\pm 4^\circ$) and $6S$---$5D$ mixing (with
$\theta=40^\circ\pm 5^\circ$) are taken into account. The
possibility to observe higher ``mixed $D$-wave" resonances,
$\tilde\Upsilon(n\,{}^3D_1)$ with $n=3,4,5$ is discussed. In
particular, $\tilde\Upsilon(\approx 11120)$, originating from the
pure $5\,{}^3D_1$ state, can acquire a rather large dielectron
width, $\sim 130$ eV, so that this resonance may become manifest
in the $e^+e^-$ experiments. On the contrary, the widths of pure
$D$-wave states are very small, $\Gamma_{ee}(n{}^3 D_1) \leq 2$
eV.

\end{abstract}

\maketitle

\section{Introduction}
\label{sect.1}

The spectrum of bottomonium  is very rich with a large number of
the levels below the $B \bar B$ threshold. Among them three well
established $\Upsilon(n\,{}^3S_1)~(n=1,2,3)$ mesons \cite{ref.1}
and the $1\,{}^3D_2$ state, discovered in \cite{ref.2}. There
exist numerous studies of low-lying bottomonium levels, where
different QCD motivated models are used
\cite{ref.3,ref.4,ref.5,ref.6,ref.7,ref.8,ref.9}. However, a
number of these levels have not been observed yet.

Observation of the $D$-wave states which lie below the open beauty
threshold, is a difficult experimental task, as demonstrated in
the CLEO experiment \cite{ref.2}, where to discover the
$1\,{}^3D_2$ level, four-photon cascade measurements in the
$\Upsilon(3S)$ radiative decays  have been performed. In
particular, neither the $1\,{}^3D_1$ state nor the members of the
$2D$ multiplet, for which  potential model calculations  give
masses around 10.45 GeV \cite{ref.4,ref.10}, i.e., below the
$B\bar B$  threshold, are as yet observed. One of the reasons for
that is the very small dielectron widths of pure $n\,{}^3D_1$
states (for any $n$): here and in \cite{ref.10} their values $\sim
1$ eV are obtained. For that reason an observation of a pure
$D$-wave vector state directly in $e^+e^-$ experiments seems to be
impossible at the present stage.

However, the situation may change for the $D$-wave vector states
which lie above the open beauty threshold. For these bottomonium
states the dielectron widths may become larger, as happens in the
charmonium family, where the experimental dielectron width of
$\psi(3770)$ (which is only 30 MeV above the $D\bar D$ threshold)
is already ten times larger than for a pure $1\,{}^3D_1$ state
\cite{ref.11,ref.12}. Moreover, the width of the $2\,{}^3D_1$
resonance $\psi(4160)$ is almost equal to that of $\psi(4040)$,
which therefore cannot be considered as a pure $3\,{}^3S_1$ state.
Such an increase of the dielectron width of a $D$-wave vector
state and at the same time a decrease of the width of an $S$-wave
state can occur if a rather large  $S$---$D$ mixing between both
states takes place \cite{ref.11,ref.12,ref.13}.

A theoretical study of the $S$---$D$ mixing between vector states
is more simple in bottomonium than in the charmonium family, since
the experimental dielectron widths are now measured for the six
states $\Upsilon(nS)~(n=1,...,6)$ \cite{ref.1}. It is also
essential that in the recent CLEO experiments the dielectron
widths of low-lying levels, $\Upsilon(nS)~(n=1,2,3)$, and their
ratios were measured with great accuracy \cite{ref.14}. These
three levels can indeed be considered as pure $S$-wave states,
because for them the $S$---$D$ mixing is possible only via tensor
forces, which give very small mixing angle (see the Appendix).
Then these pure $S$-wave states can be studied in the
single-channel approach (SCA). Here in particular, we use the
well-developed relativistic string Hamiltonian (RSH)
\cite{ref.15}. Moreover, just for these levels a comparison of
experimental and calculated dielectron widths and their ratios can
be considered as an important test of the theoretical approach and
also of the calculated wave functions (w.f.) at the origin.

There are not many theoretical studies of higher bottomonium
states \cite{ref.10,ref.16,ref.17,ref.18}. Strictly speaking, for
this task one needs to solve a many-channel problem, knowing the
interactions within all channels and between them. Unfortunately,
this program is not realized now, although some important steps in
this direction have been done recently in \cite{ref.19}, where a
theory of the interactions between the channels due to a strong
coupling to virtual (open) $B\bar B$ ($B_s\bar B_s$) channels was
developed.

In this paper, using the RSH for the low-lying states
$\Upsilon(nS)(n=1,2,3)$, we obtain a good description of the dielectron
widths and their ratios. After that we apply the same approach to
higher bottomonium states (above the $B\bar B$ threshold), where the
accuracy of our calculations is becoming worse: in particular, within
the SCA one cannot calculate the mass shift of a higher resonance,
which can occur owing to coupling to open channel(s). However, for
the dielectron widths it is most important to define the w.f. at the
origin with a good accuracy.

There exist several arguments in favor of the validity of the the SCA for
higher bottomonium states:

\begin{enumerate}

\item First, in charmonium this approximation gives the masses and
the dielectron widths of $\psi(3770)$, $\psi(4040)$, $\psi(4160)$,
and $\psi(4415)$  with a good accuracy, providing a
self-consistent description of their dielectron widths
\cite{ref.12,ref.13}.

\item Secondly, an open channel (e.g. $B\bar B$) can be considered as a
particular case of a four-quark system $Q\bar Q q\bar q$, and, as
shown in \cite{ref.20}, this open channel cannot significantly change
the w.f. at the origin of heavy quarkonia, because the magnitude of a
four-quark w.f. at the origin is two orders smaller than for a
$Q\bar Q$ meson.

\item Finally, the masses of higher bottomonium states, calculated in
SCA, appear to be rather close to the experimental values, giving a
difference between them equal to at most $50\pm 15$ MeV.

\end{enumerate}

For  pure $n\,{}^3D_1$ bottomonium states (any $n$) our
calculations give very small dielectron widths, $\leq 2$ eV, while
for higher $D$-wave vector states their dielectron widths can
increase owing to $S$---$D$ mixing through an open channel(s).
Here an important point is that the mass of a higher $n\,{}^3D_1$
state $(n\geq 3)$ appears to be only 40-50 MeV larger than that of
the $\Upsilon((n+1)S)$ state \cite{ref.4}, \cite{ref.10}, thus
increasing the probability of the $S$---$D$ mixing between these
resonances. We will define the mixing angle here in a
phenomenological way, as it was done in charmonium \cite{ref.11},
\cite{ref.13}.

Owing to the $S$---$D$ mixing the dielectron widths of mixed
$D$-wave bottomonium resonances appear to increase by two orders
of magnitude and reach $\sim 100\pm 30$ eV, while the dielectron
widths of the initially pure $n\,{}^3S_1$ resonances $(n=4,5,6)$
decrease. We also calculate the vector decay constants in
bottomonium and briefly discuss the possibility to observe ``mixed
$D$-wave" resonances in $e^+e^-$ experiments.

\section{Mass spectrum}
\label{sect.2}

The spectrum and the w.f. of the bottomonium vector states
$(L=0,2)$ are calculated with the help of the RSH and a universal
static potential from \cite{ref.21}. This Hamiltonian has been
successfully applied to light \cite{ref.22} and heavy-light mesons
\cite{ref.23,ref.24}, and also to heavy quarkonia
\cite{ref.7,ref.25,ref.26}. In bottomonium the spin-averaged
masses $M(nL)$ of the $nL$ multiplets are determined by a simpler
mass formula than for other mesons, because it does not contain
the self-energy and the string contributions which in bottomonium
are very small, $\leq 1$ MeV, and can be neglected \cite{ref.7}.
As a result, the mass $M(nL)$ just coincides with the eigenvalue
(e.v.) of the spinless Salpeter equation (SSE) or with the e.v. of
the einbein equation, derived in the so-called einbein
approximation (EA) \cite{ref.23}.

Here we use the EA, because the $nS$-wave functions, defined by
the EA equation, has an important advantage as compared to the
solutions of SSE: they are finite near the origin, while the
$nS$-w.f. of the SSE diverge for any $n$ and have to be
regularized (e.g. as in \cite{ref.4}), introducing unknown
parameters. At the same time the difference between the EA and SSE
masses is small, $\leq 15$ MeV (the e.v. in the SSE are always
smaller than the e.v. in the EA) and can be included in the
theoretical error. In Table~\ref{tab.1} the centroid masses
$M(nL)$ are given both for the SSE and the EA, calculated for the
same set of input parameters.

We do not discuss here the singlet ground state $\eta_b$, recently
discovered by BaBar \cite{ref.27}, because the hyperfine (HF)
interaction introduces extra, not well-known parameters, while our
goal here is to describe the bottomonium data, not introducing
extra parameters and using a universal potential which contains
only fundamental parameters - the QCD constant $\Lambda$ and the
string tension.

The masses of $\Upsilon(n\,{}^3S_1)$ are very close to the centroid
masses $M(nS)$ (with the exception of the ground state), because the HF
splittings are small \cite{ref.26}: for higher radial excitations the
difference between $M(n\,{}^3S_1)$ and the centroid mass is $\leq 4$ MeV
for $n\geq 3$. Moreover, for the $D$-wave multiplets the fine-structure
splittings are small \cite{ref.4} and therefore the calculated centroid
masses $M(nD)$ coincide with $M(n\,{}^3D_1)$ within the theoretical
error.

The RSH  is defined by  the expression from \cite{ref.15,ref.22}:
\begin{equation}
 H_0=\frac{\textbf{p}^2+m_b^2}{\omega}+\omega+V_B(r).
\label{eq.1}
\end{equation}

Here $m_b$ is the pole mass of the $b$ quark, for which the value
$m_b=4.832$ GeV is used. This number corresponds to the current
mass $m_b=4.235$ GeV, which coincides  with the conventional
current $b$-quark mass, equal to $4.20\pm 0.07$ GeV \cite{ref.1}.
The static potential $V_B(r)$,  defined below in Eq.(\ref{eq.5}),
contains the symbol B, which shows that this potential was derived
in background perturbation theory \cite{ref.15}.

In (\ref{eq.1}) the variable $\omega$ can be defined in two ways:
If the extremum condition is put on the Hamiltonian $H_0$,
$\omega$ is equal to the kinetic energy operator,
$\omega=\sqrt{\mathbf{p}^2+m_b^2}$. Substituting this operator
$\omega$ into $H_0$, one arrives at the well-known SSE. However,
the $S$-wave w.f. of the SSE diverge near the origin and for their
definition one needs to use a regularization procedure, in this
way introducing  several additional parameters.

Instead we prefer to use the EA, where the variable $\omega$ is
determined from another extremum condition, put on the e.v.
$M(nL)$. Then $\omega(nL)$ is not an operator anymore, but is
equal to the matrix element (m.e.) of the kinetic energy operator
and plays the role of a dynamical (constituent) quark  mass. This
constituent mass $\omega(nL)$ grows with increasing quantum
numbers and this fact appears to be very important for light and
heavy-light mesons \cite{ref.23,ref.24}, while in bottomonium the
difference between the dynamical mass $\omega_{nL}$ and the pole
mass  $m_b$ is not large, changing from $\sim 170$ MeV for the
$1S$ ground state up to $\sim 300$ MeV for higher states, like
$6S$.

In the framework of the EA the w.f. of heavy-light mesons have
been calculated and successfully applied to determine the
pseudoscalar decay constants of the $D$, $D_s$, $B$, and $B_s$
mesons, giving a good agreement with experiment \cite{ref.24}.

It is of interest to notice that in bottomonium the masses,
calculated in EA and SSE, and also in the nonrelativistic (NR)
case (where $\omega_{\rm NR}(nL)= m_b$ for all states), do not
differ much, even for higher states: such mass differences are
$\leq 40$ MeV (see below and Tables \ref{tab.1} and \ref{tab.2}).
Still, the w.f. at the origin, calculated in EA, takes into
account the relativistic corrections and gives rise to a better
agreement with the experimental dielectron widths than in the NR
approach.

In EA the masses $M(nL)$  are defined by the mass formula:
\begin{equation}
 M(nL)=\omega_{nL}+\frac{m_b^2}{\omega_{nL}}+
 E_{nL}(\omega_{nL}),
\label{eq.2}
\end{equation}
where $\omega(nL)$ and the e.v. $E_{nL}$ have to be defined
solving two self-consistent equations \cite {ref.12,ref.23},
namely
\begin{equation}
\left[\frac{\textbf{p}^2}{\omega_{nL}}+V_B(r)
\right]\varphi_{nL}(r) = E_{nL}~ \varphi_{nL}(r), \label{eq.3}
\end{equation}
and the equation
\begin{equation}
\omega^2_{nL}=m^2_b-\frac{\partial E_{nL}}{\partial \omega_{nL}}.
\label{eq.4}
\end{equation}
In (\ref{eq.3}) we use for all mesons the universal static
potential $V_B(r)$ from \cite{ref.7,ref.21}:
\begin{equation}
V_B(r) =\sigma(r)\, r - \frac43 \frac{\alpha_B(r)}{r},
\label{eq.5}
\end{equation}
with the following set of the parameters:
\begin{equation}
\begin{array}{ll}
m_b=4.832\,{\rm GeV}, & \Lambda_B(n_f=5)=0.335\ $GeV$,\\
M_B=0.95\,{\rm GeV},  & \sigma_0=0.178\ $GeV$^2. \label{eq.6}
\end{array}
\end{equation}
The QCD (vector) constant $\Lambda_B$, which determines the vector
coupling constant $\alpha_B(r)$ (see Eq.~(\ref{eq.7}) below),
depends on the number of flavors and can be expressed via the QCD
constant in the $\overline{\textrm{\textrm{MS}}}$ regularization
scheme; the connection between both constants has been established
in \cite{ref.21,ref.28}. In particular, the two-loop constant
$\Lambda_B(n_f=5)=335$ MeV in Eq.~(\ref{eq.6}) corresponds to the
two-loop $\Lambda_{\overline{\textrm{MS}}}=244$ MeV, since they
are related as $\Lambda_B(n_f=5)=1.3656~\Lambda_{\overline{\rm
MS}}$ \cite{ref.28}. However, one cannot exclude that for
low-lying bottomonium levels, like $1S$, $1P$, and $1D$, the
choice $n_f=4$, equal to the number of active flavors, might be
preferable, giving for their masses a better agreement with
experiment. Here for simplicity we take $n_f=5$ for all states,
because we are mostly interested in higher states, above the open
beauty threshold. The constant $\sigma_0$ occurs in  the
expression for the variable string tension $\sigma(r)$ given by
Eq.~(\ref{eq.8}).

\begin{table}[t]
\caption{\label{tab.1} The  spin-averaged masses $M(nL)$ (MeV) of
low-lying multiplets, calculated in nonrelativistic (NR) case, for
spinless Salpeter equation (SSE), and in einbein approximation
(EA). In all cases the parameters of $V_B(r)$ are taken from
Eq.~(\ref{eq.6})}
\begin{center}
\begin{tabular}{cllll}
  \hline  \hline
  State &~NR &~SSE &~EA &~~~Exp. \cite{ref.1} \\
  \hline
 $1S$ &~~9469  &~~9453  &~~9462  &~~9460.30$\pm$0.26~$(1^3S_1)$  \\
 $1P$ &~~9894  &~~9884  &~~9888  &~~9900.1$\pm$0.6  \\
 $2S$ &~10028  &~10010  &~10021  &~10023.3$\pm$0.3~$(2^3S_1)$  \\
 $1D$ &~10153  &~10144  &~10146  &~10161.1$\pm$1.7~$(1^2D_1)$  \\
 $2P$ &~10270  &~10256  &~10261  &~10260.0$\pm$0.6  \\
 $1F$ &~10355  &~10345  &~10347  & \quad-  \\
 $3S$ &~10379  &~10356  &~10369  &~10355.2$\pm$0.5~$(3^3S_1)$  \\
 $2D$ &~10460  &~10446  &~10450  &~\quad-  \\
 $3P$ &~10562  &~10541  &~10551  &~\quad-  \\
  \hline  \hline
\end{tabular}
\end{center}
\end{table}

The vector coupling in the coordinate space $\alpha_B(r)$ is defined
via the strong coupling  in the momentum space $\alpha_{B}(q)$
\cite{ref.21}:
\begin{eqnarray}
 \alpha_B(r) & = &
 \frac{2}{\pi}\int\limits_0^\infty dq\frac{\sin(qr)}{q}\alpha_B(q),
 \nonumber \\
 \alpha_B(q) & = & \frac{4\pi}{\beta_0t_B}\left(1-\frac{\beta_1}{\beta_0^2}
  \frac{\ln t_B}{t_B}\right)
\label{eq.7}
\end{eqnarray}
with $t_B=\frac{\ln(\mathbf{q}^2+M_B^2)}{\Lambda_B^2}$.

The solutions of Eq.~(\ref{eq.3}) are calculated here considering
two types of confining potential in (\ref{eq.5}): one with the
string tension equal to a constant, $\sigma_0=0.178$ GeV$^2$, and
the other with the string tension $\sigma(r)$ dependent on the
$Q\bar Q$ separation $r$. Such a dependence of the string tension
on $r$ appears if the creation of virtual light $q\bar q$ pairs is
taken into account, causing a flattening of the confining
potential at large distances, $\geq 1.0$ fm. This effect may
become important for bottomonium states with $R(nL)\geq 1.0$ fm,
giving a decrease of the masses (e.v.). The explicit expression of
$\sigma(r)$ is taken here from \cite{ref.29}, where it was deduced
from the analysis of radial Regge trajectories of light mesons:
\begin{equation}
 \sigma(r) =\sigma_0 (1-\gamma f(r))
\label{eq.8}
\end{equation}
with the parameters taken from \cite{ref.29}: $\gamma=0.40$,
$f(r\to 0) =0 $, $f(r\to\infty) =1.0$.

\begin{table}[ht]
\caption{\label{tab.2} The spin-averaged masses $M(nL)$ (MeV) of
higher bottomonium states in the NR case, for the SSE, and in
einbein approximation (EA) for the potential $V_B(r)$
(\ref{eq.5})}
\begin{center}
\begin{tabular}{cllll}
\hline  \hline
 State &~NR &~SSE &~EA  &~~~Exp. \cite{ref.1} \\
\hline
 $2F$ &~10623  &~10607  &~10613 &  ~\quad-  \\
 $4S$ &~10657  &~10630  &~10645 & ~10579.4$\pm$1.2~$(4^3S_1)$~  \\
 $3D$ &~10717  &~10698  &~10705 & ~\quad-  \\
 $4P$ &~10808  &~10783  &~10795 & ~\quad-  \\
 $3F$ &~10857  &~10835  &~10844 & ~\quad-  \\
 $5S$ &~10894  &~10862  &~10880 & ~10865$\pm$8~($\Upsilon(10860)$)~  \\
 $4D$ &~10942  &~10916  &~10928 & ~\quad-  \\
 $5P$ &~11024  &~10998  &~11009 & ~\quad-  \\
 $6S$ &~11100  &~11067  &~11084 & ~11019$\pm$8~($\Upsilon(11020)$)~ \\
 $5D$ &~11139  &~11109  &~11123 & ~\quad-  \\
 $7S$ &~11278  &~11240  &~11262 & ~\quad-  \\
 $6D$ &~11310  &~11270  &~11295 & ~\quad-  \\
\hline  \hline
\end{tabular}
\end{center}
\end{table}

In Tables \ref{tab.1} and \ref{tab.2} the masses $M(nL)$ are given
only for the flattening potential (\ref{eq.8}): for low-lying
levels they coincide with the masses calculated using the linear
potential (with $\sigma=\textrm{const}=\sigma_0$) within $\leq 2$
MeV. For higher states and using the flattening potential, the
masses (e.v.) are smaller by $\sim 10-60$ MeV (see the numbers in
Table~\ref{tab.3}). In particular, the mass difference is only 12
MeV for the $4S$ and $3D$ states and already 40 MeV for the $6S$
and $5D$ states, reaching 64 MeV for the $7S$ state. It is evident
that for a flattening potential the masses $M(nS)$ (any $n$) are
closer to the experimental values.

\begin{table}[t]
\caption{\label{tab.3} The masses of higher bottomonium states
(MeV) for the  static potential Eq.~(\ref{eq.5}) with the
parameters (\ref{eq.6}) and two confining potentials: linear with
$\sigma_0=0.178$ GeV$^2$ and the flattening potential
Eq.~(\ref{eq.8})}
\begin{center}
\begin{tabular}{lccccccccc}
\hline  \hline
 ~State & $2D$  & $4S$  & $3D$  & $4P$  & $5S$  & $4D$  & $6S$  & $5D$  & $7S$\\
\hline
 ~Linear~\quad &~10456~ &~10656~ &~10717~ &~10812~ &~10901~ &~10950~   &~11122~ &~11163~ &~11326~\\

 ~Flatt.\quad &10450 &10645 &10705 &10795   &10880 &10928 &11084 &11123 &11262\\
\hline  \hline
\end{tabular}
\end{center}
\end{table}

For a comparison in Tables~\ref{tab.1} and \ref{tab.2} the masses
$M_{\rm NR}(nL)$, calculated in the NR approximation (where
$\omega(nL)$=\textrm{const}=$m_b$ for all states) are also given
for the same static potential. These masses $M_{\rm NR}(nL)$ are
always 10$-$20 MeV larger than in the  EA, but in its turn the EA
masses are 10$-$20 MeV larger that the e.v. of the SSE. Such a
small difference between the EA and SSE masses is taken into
account here by including it in the theoretical error.

For our further analysis it is also important that due to the
flattening effect the w.f. at the origin (for higher states) are
becoming significantly smaller than for the linear potential, providing
a better agreement with the experimental dielectron widths.

Sometimes the point of view is taken that in bottomonium the
nonperturbative effects (determined by the confining potential)
play an insignificant role for low-lying levels. To clarify this
point we have compared two m.e. for a given $nS$ state: of the
confining (nonperturbative) potential $<\sigma(r) r>$ and of the
gluon-exchange (GE) (or ``perturbative") potential, respectively,
introducing their ratio $\eta(nS)$:
\begin{equation}
\eta(nS)=\frac{<\sigma(r) r>_{nS}}{<|V_{\textrm{GE}}(r)|>_{nS}}.
\label{eq.9}
\end{equation}
The results of our calculations are presented in
Table~\ref{tab.4}.

\begin{table}[ht]
\caption{\label{tab.4} The ratios $\eta(nS)$}
\begin{center}
\begin{tabular}{lccccccc}
\hline\hline
 ~State~      & $1S$ & $2S$ & $3S$ & $4S$ & $5S$ & $6S$ \\
\hline
 $\eta(nS)$ &~0.24~&~0.93~&~1.80~&~2.78~&~3.87~&~5.12~\\
\hline\hline
\end{tabular}
\end{center}
\label{tab.IV}
\end{table}

The values of $\eta(nS)$ from Table~\ref{tab.4} show that only for
the $1S$ ground state the nonperturbative contribution is rather
small, equal to $24\%$, while already for the $2S$ state both
contributions are equally important. For higher $nS$ states the
nonperturbative contribution dominates, being $\sim (n-1)$ times
larger than the perturbative one. For that reason the GE potential
can even be considered as a perturbation for higher resonances.

We estimate the accuracy of our calculations to be equal to 15
MeV. The calculated masses weakly depend on the admissible
variations of the parameters taken (the same accuracy was obtained
in studies of heavy-light mesons \cite{ref.24} and the charmonium
family \cite{ref.12}). Still, for higher resonances the accuracy
of the calculated masses is worse, since the influence of open
channel(s) is not taken into account. Here we can only estimate
possible hadronic (decay channel) shifts, while comparing
calculated and experimental masses: for $\Upsilon(10580)$ and
$\Upsilon(11020)$ a downward shift $\sim 50\pm15$ MeV is expected,
while the mass $M(5S)$, calculated in single-channel
approximation, is close to the experimental mass of
$\Upsilon(10860)$ (see Table~\ref{tab.2}).

Up to now, many bottomonium states, even those which lie below the
$B\bar B$ threshold, have not yet been discovered, among them the
$1D$ multiplet (two states), the $2D$ and $1F$  multiplets, and
maybe, the $3P$ multiplet, for which the centroid mass
$M(3P)=10550(15) MeV\footnote{Here and below in the brackets we
give a theoretical uncertainty}$ , very close to the threshold, is
predicted (see Table~\ref{tab.1}). The observation of these
``missing" levels would be very important for the theory.

For further analysis it is also important that the differences
between the masses of the $(n+1)S$ and $nD$ states ($n\geq 3$) are
small, decreasing for larger $n$: their values are equal to 60,
48, and 39 MeV for $n=3$, $4$, and $5$, respectively (see
Table~\ref{tab.3}).

The w.f. of the $nS$ and $nD$ states are given in the Appendix,
together with m.e. like $\omega(nL)$,
$\langle\textbf{p}^2\rangle$, and those which are needed to
determine the dielectron widths and vector decay constants. Also
we estimate the relativistic corrections, calculating the
velocities $v^2/c^2$ for different states: their values do not
change much, from 0.07 for $\Upsilon(1S)$ up to 0.11 for
$\Upsilon(6S)$ (see the Appendix). These numbers illustrate the
accuracy of the NR approximation.

In conclusion we would like to stress two points again: first, in
bottomonium the centroid masses $M(nL)$ coincide with the e.v. of
the dynamical equation; secondly, the nonperturbative dynamics
dominates for all $\Upsilon(nS)$ with $n\geq 2$.

\section{Dielectron widths}
\label{sect.3}

The dielectron widths are defined here with the help of the van
Royen$-$Weisskopf formula \cite{ref.30} and taking into account
the QCD radiative corrections \cite{ref.31}. The widths
$\Gamma_{ee}(nS)$ and $\Gamma_{ee}(nD)$ can also be expressed
through the vector decay constants $f_V$, for which explicit
expressions were derived in the framework of the field correlator
method in \cite{ref.24}. For the $S$-wave states we have
\begin{equation}
 \Gamma_{ee}(n\,{}^3S_1)=\frac{4 \pi e_b^2\alpha^2}{3M_{nS}}f_V^2(nS)\beta_V
 = \frac{4e^2_b\alpha^2}{M^2_{nS}}|R_{nS}(0)|^2\xi_{nS} \beta_V,
\label{eq.10}
\end{equation}
and a similar expression is valid for the $D$-wave vector states:
\begin{equation}
 \Gamma_{ee}(n\,{}^3D_1)=\frac{4 \pi e_b^2\alpha^2}{3M_{nD}}f_V^2(nD)\beta_V
 = \frac{4e^2_b\alpha^2}{M^2_{nD}}|R_{nD}(0)|^2\xi_{nD} \beta_V,
\label{eq.11}
\end{equation}
if the $D$-wave w.f. at the origin is defined according to the
expression Eq.~(\ref{eq.14}) below, which was derived in
\cite{ref.32}.

In Eqs.~(\ref{eq.10}) and (\ref{eq.11}) the QCD one-loop
perturbative corrections enter via the factor $\beta_V$
\cite{ref.31}:
\begin{equation}
 \beta_V=1-\frac{16}{3\pi}\alpha_s(M_V).
\label{eq.12}
\end{equation}
However, one cannot exclude that higher order perturbative
corrections may not be small and therefore, strictly speaking the
factor $\beta_V$, as well as $\alpha_s(M_V)$ in Eq.~(\ref{eq.12}),
has to be considered as an effective constant. Nevertheless this
factor cannot be used as an arbitrary parameter. In different
approaches its value typically varies in the range $0.75\pm 0.05$
\cite{ref.8,ref.16,ref.33}, which corresponds to an effective
coupling $\alpha_s(M_V)=0.14\pm 0.04$. (About the choice of the
renormalization scale, taken here equal to the mass of a vector
$b\bar b$ meson $M_V$, see the discussion in \cite{ref.34}). Here
we will neglect in the scale the difference between the mass
values for higher states, since all of them lie in narrow range,
10.6$-$11.1 GeV.

As a first step we analyse here the dielectron widths of low-lying
levels $\Gamma_{ee}(\Upsilon(nS))~(n=1,2,3)$ and their ratios
$r(m/n)$, because these do not depend on the factor $\beta_V$. As
a second step, the values of $\beta_V$ are extracted from the
magnitudes of the dielectron widths, which are now known with
great accuracy owing to the CLEO data \cite{ref.14}. Surprisingly,
just the same value $\beta_V=0.80\pm 0.01$ is extracted from our
fits to three dielectron widths $\Gamma_{ee}(nS)~(n=1,2,3)$. This
value of $\beta_V$ corresponds to $\alpha_s(\sim 10.6\,
$GeV$)=0.12\pm0.01$, which appears to be $\sim 15\%$ smaller than
the strong coupling $\alpha_s(10.330\, $GeV$)=0.142\pm 0.056$,
recently extracted from the CLEO data on the total cross sections
in $e^+e^{-}$ annihilation \cite{ref.35}.

It is reasonable to assume that such a difference may occur
due to second and third order perturbative corrections, which were
taken into account in the CLEO analysis, while second and higher-order
perturbative corrections to the dielectron widths are not calculated
yet. Taking the central value from \cite{ref.35},  $\alpha_s(10.330\,
$GeV$)=0.142$, one obtains  $\beta_V=0.76$, which is only $5\%$
smaller than our number $\beta_V(M_V)=0.80$. From  this comparison
one can estimate that in bottomonium the contribution from unknown
higher order corrections to the dielectron width is positive and
small, $\leq 10\%$.

In theoretical studies of the dielectron widths and vector decay
constants  QCD radiative corrections are often neglected, i.e.,
$\beta_V=1.0$ is taken \cite{ref.36,ref.37,ref.38}, while in our
analysis only with $\beta_V=0.80(1)$ a good description of the
dielectron widths is achieved. On the contrary, in \cite{ref.17} a
significantly smaller number, $\beta_V=0.46$, is exploited.
Probably, such a small value of $\beta_V$ (or large strong
coupling) has been used in \cite{ref.17} in order to suppress the
large values of the w.f. at the origin for low-lying states,
obtained in their model.

Thus we start with  the ratios of the dielectron widths for the
$n\,{}^3S_1$ states $(n=1,2,3)$:
\begin{equation}\label{eq.13}
 r(m/n) = \frac{\Gamma_{ee}(mS)}{\Gamma_{ee}(nS)}=
 \frac{(M(nS)^2)(R_{mS}(0))^2}{(M(mS)^2)(R_{nS}(0))^2},
\end{equation}
which are fully determined by the w.f. at the origin (the masses
are known from experiment). Taking the w.f. at the origin
calculated here, from the Appendix, one arrives at the values of
$r(m/n)$ given in Table ~\ref{tab.5}.

\begin{table}[h]
\caption{\label{tab.5} The ratios of the dielectron widths
$r(m/n)$ for low-lying $n\,{}^3S_1$ states}
\begin{center}
\begin{tabular}{llll}
\hline\hline
             &~ $r(2/1)$  &~  $r(3/1)$   &~  $r(3/2)$ \\
\hline
~Theory\quad &~ 0.465   &~ 0.339    &~ 0.728  \\
~Exp. \cite{ref.14} \quad   &~$0.457\pm0.008$&~ $0.329\pm0.006$ &~ $0.720\pm0.016$\\
\hline\hline
\end{tabular}
\end{center}
\end{table}

Both the calculated and the experimental ratios agree with each other
with an accuracy of $\leq 3\%$ and this result can be considered
as a good test of our approach, as well as of the w.f. at the origin
calculated here.

Next we calculate the absolute values of the dielectron widths,
which allow to extract the QCD factor $\beta_V$. From three
dielectron widths $\Gamma_{ee}(nS)~(n=1,2,3)$ the same value
$\beta_V=0.80(1)$ has been extracted. Later everywhere
$\beta_V=0.80$ is used, for which the dielectron widths of
low-lying and higher states are given in Tables~\ref{tab.6} and
\ref{tab.7}, respectively.

\begin{table}[h]
\caption{\label{tab.6} The dielectron widths of $n\,{}^3S_1~
(n=1,2,3)$ and $n\,{}^3D_1~(n=1,2)$ keV with $\beta_V=0.80$}
\begin{center}
\begin{tabular}{ccc}
\hline\hline
 Widths  &  Theory & Exp. \cite{ref.14} \\
\hline
 $~\Gamma_{ee}(1S)~$ & 1.320   &~1.354$\pm$0.024 \\
 $\Gamma_{ee}(2S)$ & 0.614   &~0.619$\pm$0.014 \\
 $\Gamma_{ee}(3S)$ & 0.447   &~0.446$\pm$0.011 \\ \hline
 $\Gamma_{ee}(1D)$ &~0.614$\times10^{-3}$    &  \\
 $\Gamma_{ee}(2D)$ &~1.103$\times10^{-3}$    &  \\
\hline\hline
\end{tabular}
\end{center}
\end{table}

For low-lying levels the dielectron widths (with $\beta_V=0.80$)
agree with the experimental numbers within 3\% accuracy (see
Table~\ref{tab.6}).

The dielectron widths calculated here are compared with other
theoretical predictions \cite{ref.17,ref.18} in Table~\ref{tab.7}:
in \cite {ref.17} rather small dielectron widths are obtained,
mostly due to the small $\beta_V=0.46$ taken there. This value is
$70\%$ smaller, i.e., the QCD radiative corrections are larger,
than in our case. In \cite{ref.18}, as well as in our
calculations, for $\Upsilon(10580)$ the dielectron width is larger
than in experiment, while for the ground state their dielectron
width is three times smaller than in our calculations and in
experiment.

\begin{table}[h]
\caption{\label{tab.7} The dielectron widths $\Gamma_{ee}(nS)$
(keV) of pure $S$-wave states}
\begin{center}
\begin{tabular}{ccccccc}
\hline\hline
 State        &  $1S$  & $2S$  &   $3S$  &   $ 4S$ &     $5S$ &  $6S$   \\
\hline
 GVGV\footnote{The numbers given are taken from
 second paper in \cite{ref.17}} \cite{ref.17} &  1.01  &     0.35  &    0.25  &    0.18  &   0.14&-\\
 CO \cite{ref.18}   & 0.426  &   0.356 &    0.335 &   0.311&-&-\\
 This paper &  1.320  &  0.614  &  0.447 &    0.372 &   0.314 & 0.270\\
 Exp. \cite{ref.1} &~$1.340\pm0.018$ &~$0.612\pm0.011$&~$0.443\pm0.008$&~$0.272\pm0.029$ &~$0.31\pm0.07$&~
  $0.13\pm0.03$\\
\hline\hline
\end{tabular}
\end{center}
\end{table}

In conclusion of this section we would like to stress again that:

\begin{enumerate}

\item The value $\beta_V=0.80(1)$ should be considered as an
effective constant, which implicitly takes into account the
contributions from higher perturbative corrections. We expect that
higher-order perturbative corrections are positive and rather
small, $\leq 10\%$). In the absence of higher corrections, the
effective coupling, $\alpha_s(10.6~{\rm GeV}) \sim 0.12(1)$ taken
here, appears to be slightly smaller than the strong coupling,
extracted from the analysis of the cross sections of
$e^+e^-\rightarrow \textrm{hadrons}$ \cite{ref.35}.

\item The calculated ratios of the dielectron widths (for low-lying
levels), which are independent of the unknown QCD factor $\beta_V$, agree
with experimental ratios with an accuracy better than $3\%$. Therefore
one can expect that in our approach the w.f. at the origin (for low-lying
levels) are calculated with a good accuracy.

\item The dielectron widths calculated here (with $\beta_V=0.80$)
agree with experiment with an accuracy better than $5\%$.

\end{enumerate}

\section{The S$-$D mixing between the $(n+1)\,{}^3S_1$ and $n\,{}^3D_1$
bottomonium states} \label{sect.4}

In contrast to the case of the low-lying levels, the calculated dielectron widths of pure
$nS$ vector states with $n=4$ and $n=6$ exceed the experimental values:
for the $4S$ and $6S$ states they are $25\%$ and two times larger than
the experimental widths of $\Upsilon(10580)$ and $\Upsilon(11020)$,
respectively. Such a suppression of the dielectron widths occurs
if one or more channels are open. Some reasons for that have been
discussed in \cite{ref.16}, where it was that in particular in the
Cornell coupled-channel model \cite{ref.39} the dielectron widths of
higher charmonium states are not suppressed.

Here as in \cite{ref.13}, we assume that an open channel cannot
significantly affect the w.f. at the origin calculated in
closed-channel approximation. This assumption is based on the
study of a four-quark system in \cite{ref.20}, where the
calculated w.f at the origin of a four-quark system, like $Q\bar Q
q\bar q$, appears to be about two orders smaller than that of a
heavy meson $Q\bar Q$. We expect this statement also to be true of
a continuum w.f. at the origin (the w.f. of an open channel),
which can be considered as a particular case of a four-quark
system (this does not exclude that a continuum channel can
strongly affect the $Q\bar Q$ w.f. at large distances). Thus it is
assumed that a suppression of the dielectron widths of higher
states occurs due to the $S$---$D$ mixing between the $(n+1)S$ and
$nD$ vector states, which happen to have close values of their
masses. We also show in the Appendix that in bottomonium the
$S$---$D$ mixing due to tensor forces appears to be very small,
giving a mixing angle $\theta_T < 1^\circ$.

For the $D$-wave states their w. f. at the origin is defined here
as in \cite{ref.32}:
\begin{equation}
\label{eq.14} R_D(0)= \frac{5R_D''(0)}{2\sqrt{2}\omega_b^2},
\end{equation}
and for the mixed states their physical (mixed) w.f. are given by
\begin{eqnarray}
 R_{{\rm phys}~S}(0) & = & \cos\theta R_S(0)- \sin\theta \ R_D(0),
\label{eq.15}
 \\
 R_{{\rm phys}~D}(0) & = & \sin\theta R_S(0)+ \cos\theta\ R_D(0).
\label{eq.16}
\end{eqnarray}
The w.f. at the origin of pure $S$- and $D$-wave states and the
derivatives $R_{nD}''(0)$ are given in the Appendix together with
other m.e. which are needed to calculate the physical w.f. at the
origin (see Table~\ref{tab.14}). As seen from Table~\ref{tab.12},
the w.f. $R_{nD}(0)$ are small and therefore the dielectron widths
of pure $n\,{}^3D_1$ states appear to be very small, $\leq 2$ eV.
Their values are given in Table~\ref{tab.8}. Notice that our
widths are $\sim 10$ times smaller than those in \cite{ref.36}).
On the contrary, the dielectron width of the $4\,{}^3S_1$
resonance is 25\% larger.

To obtain agreement with the experimental value,
$\Gamma_{ee}(10580)=0.273\pm 0.022$ keV, we take into account the
$4S$---$3D$ mixing  and determine the mixing angle,
$\theta=27^\circ\pm 4^\circ$ from this fit. Thus $\Upsilon(10580)$
cannot be considered a pure $4S$ vector state, it is mixed with
the initially  pure $3\,{}^3D_1$ state. This second ``mixed" state
will be denoted here as $\tilde \Upsilon(\sim 10700)$, it acquires
the dielectron width $\Gamma_{ee}(\tilde\Upsilon (10700))=0.095$
keV, which is 60 times larger than the width of the pure
$3\,{}^3D_1$ state.

The dielectron widths are given in Table~\ref{tab.8} in two cases:
without the $S$---$D$ mixing ($\theta=0$) and for mixed states,
taking the same mixing angle $\theta=27^\circ$ for all higher
states.

An interpretation of the  experimental width of $\Upsilon(10860)$ cannot
be done in an unambiguous way: the calculated $\Gamma_{ee}(5\,{}^3S_1)$
of a pure $5S$ state ($\theta=0$) just coincides with the central
value of the experimental $\Gamma_{ee}(\Upsilon(10860))=0.31\pm 0.07$
keV. It could mean that for some unknown reason the $5S$ and $4D$
vector states are not mixed. However, there exists another possibility,
because the width $\Gamma_{ee}(10860)$ has a rather large experimental
error. In particular, for the mixing angle $\theta=27^\circ$ one
obtains $\Gamma_{ee}(\Upsilon(10860))=0.23$ keV, which just coincides
with the lower bound of the experimental value. To decide which
of the two possibilities is realized, more precise measurements of
$\Gamma_{ee}(\Upsilon(10860))$ are needed.

\begin{table}[h]
\caption{\label{tab.8} The dielectron widths (keV) for the mixing
angles $\theta=0$ and $\theta=27^\circ$ ($\beta_V=0.80$)}
\begin{center}
\begin{tabular}{cccc}
\hline\hline
~Widths~   &\multicolumn{2}{c}{Theory}   & Exp. \cite{ref.14} \\
\cline{2-3}       &no mixing          &with mixing &  \\
\hline
$\Gamma_{ee}(4S)$ & 0.372             &~0.273 &~0.272$\pm$0.029~\\
$\Gamma_{ee}(3D)$ &~ 1.435$\times 10^{-3}$&0.095 &  \\
$\Gamma_{ee}(5S)$ & 0.314             &0.230 & 0.31$\pm$0.07 \\
$\Gamma_{ee}(4D)$ &~ 1.697$\times 10^{-3}$&0.084 &  \\
$\Gamma_{ee}(6S)$ & 0.270             & 0.196 & 0.13$\pm$0.03 \\
$\Gamma_{ee}(5D)$ &~ 1.878$\times 10^{-3}$& 0.075 &  \\
\hline\hline
\end{tabular}
\end{center}
\end{table}

An interesting opportunity can be realized for the originally pure
$5\,{}^3D_1$ resonance. The experimental dielectron width of
$\Upsilon(11020)$ is very small, $\Gamma_{ee}(11020)=0.13\pm 0.03$
keV, i.e., it is two times smaller than the number calculated here
without the $6S$---$5D$ mixing (i.e., $\theta=0$). Even for the
mixing angle $\theta=27^\circ$, the theoretical value is still
$26\%$ larger compared to the experimental one (see
Table~\ref{tab.8}). To fit the experimental number a rather large
mixing angle, $40^\circ\pm 5^\circ$, has to be taken. For a such a
large angle the dielectron widths of both resonances,
$\tilde\Upsilon(5D)$ (with mass $\sim 11120$ MeV) and
$\Upsilon(11020)$, appear to be almost equal:
\begin{equation}\label{eq.17}
\left\{
\begin{array}{lll}
\Gamma_{ee}(\Upsilon(11020))&=&0.137\pm 0.025~~ \textrm{keV}, \\
\Gamma_{ee}(\tilde\Upsilon(5D))&=&0.135\pm 0.025~~ \textrm{keV}.
\end{array}
\right.
\end{equation}
It is of interest to notice that this large angle is close to the
value of the mixing angle $\theta\cong 35^\circ$, which has been
extracted in \cite{ref.12} and \cite {ref.40} to fit the dielectron widths in
the charmonium family: $\psi(4040)$, $\psi(4160)$, and
$\psi(4415)$.

\section{Decay constants in vector channels}
\label{sect.5}

The decay constant in vector channel $f_V(nL)$ is expressed via
the dielectron width in a simple way, as in Eqs.~(\ref{eq.10}) and
(\ref{eq.11}). Therefore, from the experimental widths the
``experimental" decay constants can be easily obtained. Still an
uncertainty is left, coming from the theoretical error of about
$10\%$ in the QCD factor $\beta_V$. Also in many papers
perturbative one-loop corrections are neglected, i.e.,
$\beta_V=1.0$ is taken \cite{ref.36,ref.37,ref.38}. This makes a
comparison with other calculations more difficult. In our
calculations we take $\beta_V=0.80$, which is slightly more than
$\beta_V\sim 0.70$, used in \cite{ref.8} and \cite{ref.33}.

To determine the experimental $f_V(\textrm{exp})$ we take in this
section the experimental data from PDG \cite{ref.1} (not the CLEO
data \cite{ref.14}), which are used in most of the cited
theoretical papers. These decay constants are given in
Table~\ref{tab.9}, both for $\beta_V=1.0$ and $\beta_V=0.80$, the
difference between them is $\sim 10\%$. The theoretical
predictions for $f_V$ give significantly different numbers
\cite{ref.36,ref.37,ref.38}, which shows that the decay constants
are rather sensitive to the dynamical parameters of the
interaction and the model used. For a comparison we give in Tables
~\ref{tab.9} and \ref{tab.10} the decay constants $f_V(nS)$ and
$f_V(nD)$, calculated here and in Ref.~\cite{ref.36}, where the
relativistic Bethe$-$Salpeter method was used. All values needed
for our calculations are presented in the Appendix.

\begin{table}
\caption{ \label{tab.9} The decay constants $f_V(nS)$ (MeV)}
\begin{center}
\begin{tabular}{ccccccc}
\hline\hline
State  &~~~~$1S$~~~~      &~~~~$2S$~~~~    &~~~~$3S$~~~~  &~~~~$4S$~~~~ &~~~~$5S$~~~~ &~~~~$6S$~~~~  \\
\hline
$\beta_V=1.0$ \cite{ref.36}  &498(20) &366(27) &304(27)&259(22) &228(16) &- \\
~This paper, $\beta_V=0.80$&794  &557 &483 &383 &355 &331\\
~$f_V$(exp) for $\beta_V=0.80$  \cite{ref.1}   &$798\pm6$&
$556\pm6$& $481\pm5$& $381\pm19$& $413\pm45$&
 $268\pm30$\\
$f_V$(exp) for $\beta_V=1.0$  \cite{ref.1}  &$715\pm5$& $497\pm5$&
$430\pm4$&  $341\pm17$& $369\pm40$&
 $240\pm27$\\
\hline\hline
\end{tabular}
\end{center}
\end{table}

For a comparison we also mention here the values of $f_V(1S)$ from
\cite{ref.37,ref.38} where the QCD factor $\beta_V=1.0$ was used:
$f_V(1S)= 529$ MeV in \cite{ref.37} is significantly smaller than
the value $f_V(1S)=705(27)$ MeV in \cite{ref.38}, which is very
close to the ``experimental'' $f_V(1S)$ (see Table~\ref{tab.9}).
On the contrary, in \cite{ref.33} the perturbative corrections
have been taken into account with $\beta_V\sim 0.66$. There the
values of $f_V$ are not given but the dielectron widths of the
$nS$ vector states $(n=1,2,3)$ are in reasonable agreement with
experiment.

For the $D$-wave states the w.f at the origin, the second
derivatives, and other m.e. determining the vector decay constants
$f_V(nD)$, are given in the Appendix. The calculated $f_V(nD)$ are
presented in Table~\ref{tab.10} together with the numbers from
\cite{ref.36}. Unfortunately, at present there are no experimental
data on the dielectron widths for those states.

\begin{table}
\caption{\label{tab.10} The decay constants $f_V(nD)$ (MeV)}
\begin{center}
\begin{tabular}{cccccc}
\hline\hline
 State &  ~~~~$1D$~~~~ & ~~~~$2D$~~~~& ~~~~$3D$~~~~  &  ~~~~$4D$~~~~ & ~~~~$ 5D$~~~~\\
\hline
 $\beta_V=1.0$~\cite{ref.36} & 261(21) & 155(11)& 178(10)&-&-\\
 This paper, $\beta_V=1.0$,
 $\theta=0^\circ$       &18   &  24   &   28    &~31~ &~33~\\
 ~~~This paper, $\beta_V=1.0$, $\theta=27^\circ$~\footnote{In bottomonium the
$2S-1D$ and $3S-2D$ states, occuring below the threshold, do not mix via
tensor forces, see a discussion in Appendix.}
 & -   & -     & 226    &~215~&~206~\\
\hline\hline
\end{tabular}
\end{center}
\end{table}

The decay constants of pure $nD$ vector states $(n=1,2,3)$, calculated in
our approach, appear to be  $\sim 10$ times smaller than those from
\cite{ref.36}, where the Bethe$-$Salpeter equation was used, and the
reason behind such a large discrepancy remains unclear. However, if
the $4S$---$3D$ mixing is taken into account (with $\theta=27^\circ$),
then the values of $f_V(3D)$ are close to each other in both approaches.

\section{Summary and Conclusions}
\label{sect.6}

In this paper we have calculated the bottomonium spectrum and
shown that the masses of the $(n+1)S$ and $nD$ states (for a given
$n\geq 2$) are close to each other. We also assume here that
between these states $S$---$D$ mixing takes place, which allows to
describe the dielectron widths of higher states with a good
accuracy. There are several arguments in favor of such a mixing.

\begin{enumerate}

\item Suppression of the dielectron widths of $\Upsilon(10580)$ and
$\Upsilon(11020)$.

\item Similarity with the $S$---$D$ mixing effects in the charmonium family.

\item Strong coupling to the $B\bar B$ $ (B_s\bar B_s)$ channel. This
fact has been supported by recent observations of the resonances
in the processes like $e^+e^-\rightarrow \Upsilon(nS)
\pi^+ \pi^-~ (n=1,2,3)$  and the theoretical analysis in \cite{ref.19}.
\end{enumerate}

The important question arises whether it is possible to observe
the mixed $D$-wave vector resonances in $e^+e^-$ experiments. Our
calculations give $M(3D)=10700(15)$ MeV (not including a possible
hadronic shift) and $\Gamma_{ee}(\tilde\Upsilon(3D))\sim 95$ eV
for the mixing angle $\theta=27^\circ$, which is three times
smaller than $\Gamma_{ee}(\Upsilon(10580))$. For such a width an
enhancement of this resonance in the $e^+e^-$ processes might be
suppressed, as compared to the peak from the $\Upsilon(10580)$
resonance.

The situation remains unclear with the $5S$---$4D$ mixing, because
the dielectron width of $\Upsilon(10860)$ contains a rather large
experimental error and a definite conclusion about the value of
the mixing angle, or no mixing at all, cannot be drawn. We have
considered both cases here, obtaining the mass $10930\pm 15({\rm
th})$ MeV for the $4D$ state.

It looks  more probable to observe the resonance
$\tilde\Upsilon(5D)$ (with the mass $\sim 11120$ MeV), for which
the  dielectron width may be almost equal to that of the
conventional $\Upsilon(11020)$ resonance. However, since the cross
sections of $e^+e^-$ processes depend also on other unknown
parameters, like the total width and branching ratio to hadronic
channels, the possibility to observe a mixed $5D$ vector resonance
may be smaller than for $\Upsilon(11020)$, even for equal
dielectron widths.

Recently new observations in the mass region 10.6--11.0 GeV have
been reported \cite{ref.41,ref.42}.  The resonance
$\Upsilon(10890)$, considered to be identical to
$\Upsilon(10860)$, has been observed by the Belle Collaboration
\cite{ref.41}. Two resonances in the same region,
$\Upsilon(10876)$ and $\Upsilon(10996)$, have been measured by the
BaBar Collaboration \cite{ref.42}, they are supposed to be the
conventional $\Upsilon(10860)$ and $\Upsilon(11020)$.  Still there
are some differences between the masses and total widths of the
resonances from \cite{ref.41,ref.42}, and the PDG data
\cite{ref.1}, so that further analysis of their parameters is
needed. Also one cannot exclude that an overlap of
$\Upsilon(11020)$ with the still unobserved $\tilde\Upsilon(5D)$
resonance is possible, which can distort the shape and other
resonance parameters of the conventional $\Upsilon(11020)$
resonance.

\begin{acknowledgments}
This work is supported by the Grant NSh-4961.2008.2. One of the
authors (I.V.D.) is also supported by the grant of the {\it
Dynasty Foundation} and the \textit{Russian Science Support
Foundation}.
\end{acknowledgments}

\section*{Appendix \\The wave functions at the origin and some
matrix elements} \label{sect.A}
\appendix

\setcounter{equation}{0}
\def\theequation{A.\arabic{equation}}

We start with the definition of the vector decay constants $f_V$,
for which the following  expression was derived  in the framework
of the vacuum correlator method in \cite{ref.24}:
\begin{equation}
\label{eq.18}
 f^2_V(nL)=12\frac{\left|\psi_{nS}(0)\right|^2}{M_V(nL)}\,
 \xi_V(nL)= \frac{3}{\pi}\frac{\left|R_{nS}(0)\right|^2}{M_V(nS)}\,
 \xi_V(nL).
\end{equation}
Besides the w.f. at the origin, the vector decay constant contains
a relativistic factor $\xi_V(nL)$, which was also defined in
\cite{ref.24}:
\begin{equation}
\label{eq.19}
 \xi_V=\frac{m^2+\omega^2+\frac{1}{3}\langle\textbf{p}^2\rangle}{2\omega^2}.
\end{equation}

Numerically, $\xi_V(nL)$ is close to unity for all $nS$ and $nD$
vector states. Moreover, for the $(n+1)S$ and $ nD$ states, which
we are mostly interested in here, their values coincide (see
Table~\ref{tab.11}). Notice that all $\xi_V(nL)$ differ from unity
at most by 4\%.

\begin{table}[h]
\caption{\label{tab.11} Relativistic factors $\xi_V(nL)$}
\begin{tabular}{cccccccccccc}
\hline\hline State & $1S$ & $2S$ & $3S$ & $4S$ & $5S$ & $6S$ &
$1D$ & $2D$ & $3D$ & $4D$ & $5D$\\ \hline
$\xi_V$ &~0.976~ &~0.974~ &~0.972~ &~0.968~ &~0.966~ &~0.963~ &~0.975~ &~0.972~ &~0.968~ &~0.966~ &~0.965~ \\
\hline\hline
\end{tabular}
\end{table}

The meaning of the m.e. $\omega_{nS}$ and $\omega_{nD}$ has been
discussed in Section 1 and their values are given in
Table~\ref{tab.12} together with the m.e. $\langle\bm{p}^2\rangle$
and the w.f. at the origin $R_{nS}(0)$ and $R_{nD}(0)$. Notice
that again the values of $\omega_{nL}$ coincide for $(n+1)S$ and
$nD$ states. Knowledge of the m.e. $\langle\textbf{p}^2\rangle$
and $\omega(nL)$ allows to define the velocities $v^2/c^2$ for
different states, which in all cases appear to be rather small,
$\leq 0.11$.

\begin{table}[h]
\caption{\label{tab.12} The matrix elements $\omega_{nS}$ (GeV),
~$\langle\textbf{p}^2\rangle$ (GeV$^2$), and the w.f. at the
origin $R_{nS}(0)$ (GeV$^{3/2}$) (no mixing) for the potential
(\ref{eq.5})}
\begin{center}
\begin{tabular}{cccccccccccc}
\hline\hline State & ~~~$1S$~~~ & ~~~$2S$~~~ & ~~~$3S$~~~ &
~~~$4S$~~~ & ~~~$5S$~~~ & ~~~$6S$~~~ & ~~~$1D$~~~ & ~~~$2D$~~~ &
~~~$3D$~~~ &
~~~$4D$~~~ & ~~~$5D$~~~ \\
\hline
$\omega_{}$ &5.02  &5.03  &5.05  &5.08  &5.1   &5.12  &5.02  &5.05  &5.08  &5.1   &5.11 \\
$R_{}(0)$   &2.529 &1.829 &1.616 &1.506 &1.424 &1.349 &0.0595&0.081 &0.0956&0.107 &0.1145\\
$\langle\textbf{p}^2\rangle$ &1.873 &1.907 &2.178 &2.459 &2.689 &2.844 &1.821 &2.136 &2.423 &2.661 &2.810\\
\hline\hline
\end{tabular}
\end{center}
\end{table}

The w.f.  $R_{nD}(0)$ are defined via the derivative
$R_{nD}^{''}(0)$ according to the definition (\ref{eq.14}) from
Ref.~\cite{ref.32}. Our calculations give the following numbers of
the derivatives, presented in Table~\ref{tab.13}.

\begin{table}[h]
\caption{\label{tab.13}The second derivative $R_{nD}^{''}(0)$
(GeV$^{7/2}$)}
\begin{center}
\begin{tabular}{cccccc}
  \hline\hline
  state            &1D    &2D    &3D    &4D    &5D     \\ \hline
  $R_{nD}^{''}(0)$ &~0.848~ &~1.167~ &~1.396~ &~1.572~ &~1.692~  \\
  \hline\hline
\end{tabular}
\end{center}
\end{table}

An interesting feature of the $D$-wave w.f. is that in contrast to
the w.f. $R_{nS}(0)$, which decrease for larger $n$, the
derivatives $R_{nD}^{''}(0)$ and  $R_{nD}(0)$ grow for higher
radial number $n$ (see  Tables ~\ref{tab.12} and \ref{tab.13}),
and this effect increases the probability of the $S$---$D$ mixing.

In bottomonium the w.f. at the origin  has been calculated in many
papers \cite{ref.16,ref.18,ref.43}. Calculated in our approach
w.f. for the pure $nS$ and $nD$ states, which  are presented in
Table~\ref{tab.12}, allow to determine the physical w.f. at the
origin of mixed states for an arbitrary mixing angle. For higher
states such physical w.f. at the origin are given in
Table~\ref{tab.14}.

For $\Upsilon(10580)$ and $\Upsilon(10860)$ we take the mixing
angle $\theta=27^\circ$, which was extracted to fit the dielectron
width $\Gamma_{ee}(10580)$. For $\Upsilon(11020)$ we take a larger
mixing angle, $\theta=40^\circ$, to fit its dielectron width.

\begin{table}[h]
\caption{\label{tab.14} The physical wave functions at the origin
$R_{\rm phys}(0)$ (GeV$^{3/2}$)}
\begin{center}
\begin{tabular}{lccc}
\hline\hline
 Mixing angle &\multicolumn{2}{c}{$\theta=27^\circ$}   & $\theta=40^\circ$ \\
\hline
  State  &~$\Upsilon(10580)$~      &  ~$\Upsilon(10860)$~  & ~$\Upsilon(11020)$~\\
 $R_{\rm phys}(0)$      &1.298      & 1.220     &   0.960 \\ \hline
 State &$\tilde\Upsilon(10700)$&$\tilde\Upsilon(10930)$&$\tilde\Upsilon(11120)$
\\
 $R_{\rm phys}(0)$   &0.769    &   0.742   &      0.955\\
\hline\hline
\end{tabular}
\end{center}
\end{table}

Just the numbers from Table~\ref{tab.14} have been used in
Sections 4 and 5 to calculate the dielectron widths and the vector
decay constants.

In this paper we do not discuss the pseudoscalar decay constants
$f_P(nS)$ in bottomonium, although for them the expressions
similar to (\ref{eq.18}) have been obtained in \cite{ref.24};
these constants need a special consideration.

In conclusion we also calculate the m.e. over the tensor forces,
which are defined by the potential $V_T(r)$:
\begin{equation}
 V_T(r) = \frac{4 \alpha_s(m_b)}{3  \omega_b^2}\frac{1}{r^3}.
\label{eq.20}
\end{equation}

Notice, that in denominator instead of $m_b^2$, usually used in
different models, in the field correlator method the tensor
potential (\ref{eq.20}) has to contain the squared dynamical mass
$\omega(nL)^2$ \cite{ref.15, ref.44}.  The $S$---$D$ mixing
between $(n+1)S$ and $nD$ vector states, due to tensor
interaction, is proportional to the nondiagonal m.e. $\langle
r^{-3}\rangle_{(n+1)S,nD}$, for which the following numbers were
obtained here (see Table~\ref{tab.15}).

\begin{table}[h]
\caption{\label{tab.15} The matrix elements $\langle
r^{-3}\rangle$ (GeV$^3$)}
\begin{center}
\begin{tabular}{cccccc}
\hline\hline
  $n$  &              1  &       2  &      3  &   4  &     5 \\
   \hline
 $\langle r^{-3}\rangle_{(n+1)S,nD}$
&  ~0.0574~&   ~0.0683~  &~0.0721~ & ~0.0739~ & ~0.0744~ \\
\hline\hline
\end{tabular}
\end{center}
\end{table}

With the help of these numbers one can easily calculate the m.e.
over the tensor potential (\ref{eq.20}), considering it as a
perturbation. These m.e. are indeed very small, $\leq 1$ MeV and
therefore give rise to small contributions to the w.f. at the
origin. These small correction to the w.f. at the origin define
the mixing angles due to tensor interaction, $\theta_T$, for
different states. We obtain that an admissible mixing angle
$\theta_T$ is small, $\theta_T\leq 1^\circ$.

The vector decay constants were calculated with the use of
Eq.~(\ref{eq.18}) and presented in Table~\ref{tab.16}.  For higher
states their values are given in two cases: without mixing and for
$\theta=27^\circ$. One can see that for the originally pure $4S$
and $6S$ states the constants $f_V$ decrease by 15\% and 35\%,
respectively, compared to the case without mixing. For originally
pure $D$-wave states their decay constants $f_V$ are growing,
namely, $f_V(3D)$ and $f_V(5D)$ increase eightfold.

\begin{table}[h]
\caption{\label{tab.16} The decay constants $f_V$ (MeV) for pure
and mixed states with $\theta=27^\circ$}
\begin{center}
\begin{tabular}{cccccccccccc}
\hline\hline
 State  & ~~$1S$~~ & ~~$2S$~~ &  ~~$3S$~~ &~~$4S$~~ & ~~$5S$~~& ~~$6S$~~&~~$1D$~~ & ~~$2D$~~ &  ~~$3D$~~  & ~~$4D$~~  & ~~$5D$~~\\
\hline
no mixing  &794 &557 &483  &444 &415&389&18 &24 &28   &31   &33 \\
 $\theta=27^\circ$ &    &    &     &  383  & 355   &331& & &226 &215 &206 \\
\hline\hline
\end{tabular}
\end{center}
\end{table}
In Table~\ref{tab.16} for low-lying mixed states the values of
$f_V(nS)$ (n\,=\,1,\,2,\,3) and $f_V(nD)$ (n\,=\,1,\,2) are
absent, because these states are indeed pure states, since for
them the $S$---$D$ mixing can occur only via tensor forces, which
give very small effect $\theta_T\leq 1^\circ$.

\end{document}